\documentclass[10pt,preprint]{aastex}
\begin{document}

%

\title{Light Propagation in Inhomogeneous Universes.\\
V. Gravitational Lensing of Distant Supernovae.}

\author{Hugo Martel\altaffilmark{1} and Premana Premadi\altaffilmark{2}}

\altaffiltext{1}{D\'epartement de physique, de g\'enie physique et d'optique,
Universit\'e Laval, Qu\'ebec, Canada, G1K 7P4}

\altaffiltext{2}{Department of Astronomy and Bosscha Observatory,
Bandung Institute of Technology, Bandung, Indonesia}

\begin{abstract}
We use a series of ray-tracing experiments to determine the
magnification distribution of high-redshift sources by gravitational
lensing. We determine empirically the relation between magnification
and redshift, for various cosmological models. We then use this
relation to estimate the effect of lensing on the determination of the
cosmological parameters from observations of high-$z$ supernovae. We
found that, for supernovae at redshifts $z<1.8$, the effect of lensing
is negligible compared to the intrinsic uncertainty in the
measurements. Using mock data in the range $1.8<z<8$, we show that the
effect of lensing can become significant. Hence, if a population of
very-high-$z$ supernovae was ever discovered, it would be crucial to fully
understand the effect of lensing, before these SNe could be used to
constrain cosmological models. 
We show that the distance moduli
$m-M$ for an open CDM universe and a $\Lambda$CDM universe are comparable
at $z>2$. Therefore if supernovae up to these redshifts were ever discovered,
it is still the ones in the range $0.3<z<1$ that would
distinguish these two models.
\end{abstract}


\newpage

%

\section{INTRODUCTION}

High-redshift supernovae have become a major tool in modern
cosmology. By measuring their apparent magnitudes, we can
estimate their luminosity distances $d_L$
(see \citealt{tonryetal03,barrisetal04,riessetal04}, and references therein).
Since the relationship
between $d_L$ and the redshift $z$ depends on the cosmological parameters,
observations of distant SNe can constrain the cosmological model. Prior
to the announcement of the {\sl WMAP} results \citep{bennettetal03},
observations of high-$z$ SNe provided the most compelling evidence of the
existence of a nonzero cosmological constant. Since then, they have been
used in combination with the {\sl WMAP} data to refine the determination
of the cosmological parameters.

The luminosity distances $d_L$ are determined by combining the observed
fluxes $F$ with estimates of the SNe luminosities $L$. Uncertainties
in $d_L$ are caused by uncertainties in $L$, because
SNe are not perfect standard candles. The flux $F$ is much easier to measure,
but for distant sources the value of $F$ might be altered by gravitational
lensing caused by the intervening distribution of matter. For instance,
a magnification $\mu>1$ would result in a increase in $F$, and an
underestimation of $d_L$.

Estimating the effect of lensing on the statistics of high-$z$ supernovae
is a complex problem. Using either an analytical model or ray-tracing
simulations, we can estimate the effect of lensing of a large
number of sources in a statistical sense. We would then need to redo
the error analysis on the SNe data to include in a consistent way the
effect of lensing. This would be a very complex task, and 
in this paper we have chosen a much simpler approach. 
{\it Our goal is not to obtain a precise estimate
of the error introduced by lensing, but rather to assess the importance
of this effect: is it dominant, important, or negligible, and for what
range of redshift? and how does it affect the discrimination between
different cosmological models?\/} To answer these questions, we take at
face value the published results of Type~Ia SNe, including their error bars
which account for every source of uncertainty but gravitational lensing.
Then, we include {\it a posteriori\/} the effect of lensing, to estimate
the change in the errors. This approach is not rigorous at all, and does
not constitute a substitute for a rigorous treatment of the errors.
But it has the great advantage of simplicity. We do not have to redo
the detailed error analysis performed by the high-redshift SNe groups,
and, more importantly, our conclusions will not be tied to any particular
sample or particular data reduction and error analysis technique
used by any particular group. We are seeking to make generic statements about
the importance of lensing (or lack of) that are relevant to any 
current or future sample of high-$z$ SNe.

The lensing of distant supernovae has been the focus of several recent 
studies. In an early study, \citet{wambsganssetal97} used ray-tracing
experiments to estimate the effect of weak lensing on the determination
of the deceleration parameter $q_0$.
\citet{md05}, \citet{dv05}, and \citet{mv06} focused on SNe as a mean
to study the nature of weak lensing.
The issue of determining the cosmological parameters for distant SNe,
and how this determination is affected by lensing,
was addressed by \citet{wang05} who used 
semi-analytical models to determine the magnification
distribution function, \citet{hl05} who used Monte Carlo
ray-tracing simulations to study the effect of weak and strong lensing,
and \citet{gunnarssonetal06} and \citet{jonssonetal06}, who
estimated the effect on lensing along individual lines of sight by considering
the properties of foreground galaxies in the same direction.
These various studies
concluded that the effect of lensing on current determinations of
the cosmological parameters is small. \citet{alderingetal06} discussed
the effect of gravitational lensing on a population of SNe at $z>1.7$.

What distinguishes our approach is mostly its
simplicity. Our calculations depend on very few assumptions, and this
implies a certain amount of robustness to our results. Even though we
rely on numerical simulations, this work should be regarded as a
back-of-the-envelope calculation, whose purpose is to obtain a qualitative
estimate of the effect of lensing on the determination of cosmological
parameters by distant SNe. Using ray-tracing experiments, rather than
a semi-analytical approach, enables us to extend our study to redshifts
much higher than the ones considered by \citet{wang05} and \citet{hl05}.

This paper is organized as follow: In \S2, we describe our calculation
of the magnification distribution $P(\mu)$, and how to estimate that
distribution at any redshift $z$. In \S3, we describe the real and mock
samples of supernovae we use for our calculations. Results are presented
in \S4. In \S5, we address various observational issues. Summary and 
conclusion are presented in \S6.

\section{THE MAGNIFICATION DISTRIBUTION FUNCTION}

\subsection{Simulations}

We have developed a {\it multiple lens-plane algorithm} to study
light propagation in inhomogeneous universes
\citep{pmm98,mpm00,premadietal01a,premadietal01b}. In this algorithm,
the space between
the observer and the sources is divided into a series of cubic boxes of
comoving size $128\,\rm Mpc$, and the
matter content of each box is projected onto
a plane normal to the line of sight. The trajectories of light rays
are then computed by adding successively the deflections caused by each plane.

To use this algorithm, we need to provide a description of the matter
distribution along the line of sight. Matter is divided into two
components: background matter and galaxies. We use a $\rm P^3M$ algorithm
\citep{he81}
to simulate the distribution of background matter. The simulations used $64^3$
equal-mass particles and a $128^3$ PM grid, inside a comoving volume of
size $128\,\rm Mpc$. The matter distribution
in the different cubes along the line of sight
then corresponds to the state of the simulation at different
redshifts.\footnote{In practice, we combine cubes from different
simulations in order to avoid periodicities along the line of sight.
See also the interesting alternative suggested by \citet{vw03}.}
We then use a Monte Carlo method for locating galaxies
into the computational volume \citep{mpm98,pmm98}. Galaxies are located
according to the underlying distribution of background matter. Morphological
types are ascribed according to the morphology-density relation
\citep{dressler80}. Galaxies are modeled as nonsingular isothermal spheres,
with rotation velocities and core radii that vary with luminosity
and morphological types.

We consider three Cold Dark Matter (CDM)
cosmological models: (1) a flat, cosmological constant model ($\Lambda$CDM)
with $\Omega_0=0.27$, $\lambda_0=0.73$, and $H_0=71\,\rm km\,s^{-1}Mpc^{-1}$.
This model is in agreement with the results of the {\sl WMAP} satellite
\citep{bennettetal03}. (2) a low-density model with 
$\Omega_0=0.3$, $\lambda_0=0$, and $H_0=75\,\rm km\,s^{-1}Mpc^{-1}$.
(3) a matter-dominated model with
$\Omega_0=1$, $\lambda_0=0$, and $H_0=75\,\rm km\,s^{-1}Mpc^{-1}$.
For each model, we consider sources at 8 different redshifts:
$z_s\simeq1$, 2, 3, 4, 5, 6, 7, and 8.\footnote{The exact values of
the source redshifts depend on the locations of the lens planes, which
vary among models.}
For each combination model-redshift, we performed
10--20 ray tracing experiments. Each experiment consists of propagating a
square beam of $101\times101=10,201$ rays back in time from the present to
redshift $z_s$, through the matter distribution. The rays in the beam were
widely separated, by 6 arc minutes,
and therefore sampled different regions of space.
We computed the magnification matrix ${\bf A}$ along each ray.
The magnification $\mu$ is then given by
\begin{equation}
\mu={1\over{\rm det}\,{\bf A}}\,.
\end{equation}

\begin{figure}
\vskip-0.3truein
\begin{center}
\includegraphics[width=4.6in]{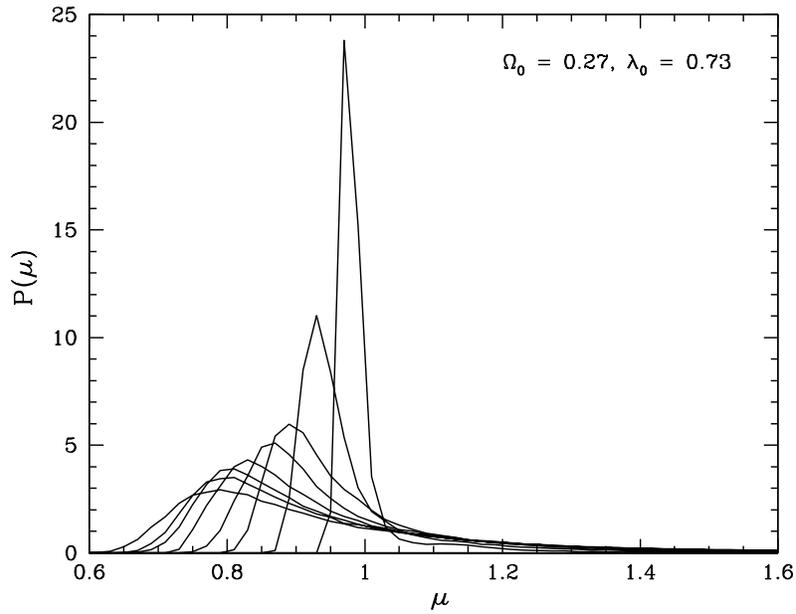}
\vskip-0.2truein
\caption{Distribution of magnifications for the
$\Lambda$CDM model. The various curves
correspond to different source redshifts: $z_s=1$ (narrowest curve),
2, 3, 4, 5, 6, 7, and 8 (widest curve).
}
\label{mu_lambda}
\end{center}
\end{figure}

\noindent
Figure~\ref{mu_lambda}
shows the distribution of magnifications for the $\Lambda$-model.
The distribution peaks at $\mu<1$, and is strongly skewed. The width of
the distribution increases with the source redshift. The distributions
for the other two models are qualitatively similar.

\subsection{Standard Deviation and Magnification Distribution}

\begin{figure}
\vskip-0.2truein
\begin{center}
\includegraphics[width=4.4in]{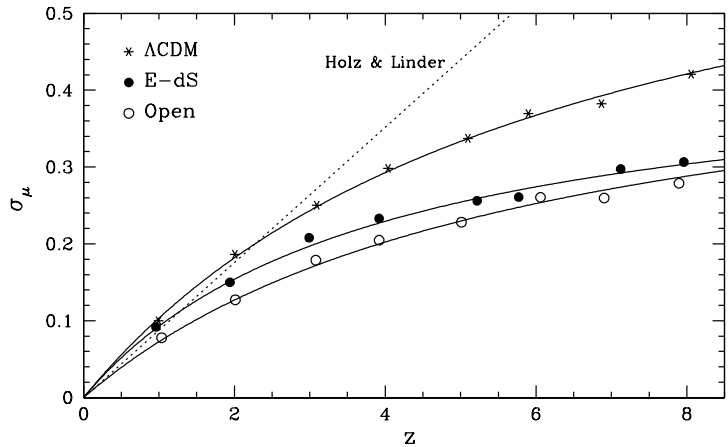}
\vskip-0.3truein
\caption{Standard deviation $\sigma_\mu$ versus redshift, for all
three models considered. The solid lines show empirical fits of the
form $\sigma_\mu=bz/(1+cz)$. The dotted line shows the relation
derived by \citet{hl05}.
}
\label{sigma}
\end{center}
\end{figure}

We have determined the distributions $P(\mu)$ at some particular
redshifts $z_s$. Since SNe do not cooperate by going off only at
these redshifts, we now want to interpolate
between these distributions, to obtain $P(\mu)$ at any redshift.
First, for each model and each source redshift $z_s$ we considered, we
compute the standard deviation $\sigma_\mu$ of the magnification distribution
$P(\mu)$. The values are shown in Figure~\ref{sigma}.
We use an empirical fit of the form
\begin{equation}
\label{sigmafit}
\sigma_\mu={bz\over1+cz}\,,
\end{equation}

\begin{deluxetable}{lccccc}
\tabletypesize{\footnotesize}
\tablecaption{Coefficients of Approximation for $\sigma_\mu$}
\tablewidth{0pt}
\tablehead{\colhead{Model} & \colhead{$\Omega_0$} & \colhead{$\lambda_0$}
& \colhead{$H_0[{\rm km\,s^{-1}\,Mpc^{-1}}]$} & \colhead{$b$} & \colhead{$c$}
}
\startdata 
$\Lambda$CDM & 0.27 & 0.73 & 71 & 0.120 & 0.16 \\
Open         & 0.30 & 0.00 & 75 & 0.085 & 0.17 \\
EdS          & 1.00 & 0.00 & 75 & 0.117 & 0.26 \\
\enddata
\end{deluxetable}

\noindent where the values of $b$ and $c$ are given in Table~1.
This enables us to estimate the values of
$\sigma_\mu$ at any redshift. Using the stochastic
universe method (SUM) of \citet{hw98},
\citet{hl05} derived a linear relation between $\sigma_{\rm eff}$ and $z$ in
the range $0\leq z\leq2$, for a $\Lambda$CDM model, where $\sigma_{\rm eff}$
is the effective standard deviation of a single measurement, which is not
the same thing as our standard deviation $\sigma_\mu$.
We plotted their result in Figure~\ref{sigma}
for comparison. There is a fairly good agreement between the two methods
at redshifts $z\leq2$. The linear relation has a slope of $0.088$.
Our empirical fit for the $\Lambda$CDM model\footnote{We use $\Omega_0=0.27$;
\citet{hl05} used $\Omega_0=0.28$.} has a slope that varies from 0.120 to
0.069 in the range $z=0-2$.

To determine $P(\mu)$ at any redshift $z$, we interpolate between the 
distributions we have already determined. Consider two known distributions
$P_1(\mu)$ and $P_2(\mu)$ at redshifts $z_1$ and $z_2$ that bracket
$z$. These distributions satisfy the properties
\begin{eqnarray}
\label{norm}
&&\int_0^\infty P_i(\mu)d\mu=1\,,\\
&&\int_0^\infty\mu P_i(\mu)d\mu=1\,,\\
\label{sigmadef}
&&\int_0^\infty(\mu-1)^2P_i(\mu)d\mu=(\sigma_\mu^2)_i\,,
\end{eqnarray}

\noindent where $i=1,2$. We define a new distribution,
\begin{equation}
\label{newpmu}
P(\mu)={[(\sigma_\mu^2)^{\phantom2}_2-\sigma_\mu^2]P_1(\mu)
+[\sigma_\mu^2-(\sigma_\mu^2)^{\phantom2}_1]P_2(\mu)\over
(\sigma_\mu^2)^{\phantom2}_2-(\sigma_\mu^2)^{\phantom2}_1}\,.
\end{equation}

\noindent We can easily check that this distribution also satisfies
the relations~(\ref{norm})--(\ref{sigmadef}). This enables us to
estimate the magnification distribution $P(\mu)$ at any
redshift $z$. We first determine
$\sigma_\mu(z)$ from equation~(\ref{sigmafit}), and then substitute
it in equation~(\ref{newpmu}) to get $P(\mu)$ at that redshift.

\noindent 

\section{THE SUPERNOVAE CATALOGS}

\subsection{The Tonry et al. Sample}

Observations of high-redshift supernovae provide an estimate of the
luminosity distance $d_L$. These results are reported in various form
in the literature. Some authors express their measurements in terms
of effective magnitudes of
distance moduli. The High-z Supernova Search Team
\citep{tonryetal03,barrisetal04,riessetal04} express their measurements 
in the following form,
\begin{equation}
\label{apmda}
\log(d_LH_0)=a\pm\delta_a\,,
\end{equation}

\noindent where $H_0$ is the Hubble constant, $a$ is the ``measurement,''
and $\delta_a$ is the ``intrinsic uncertainty,'' which includes every
possible source of error, {\it except}\/ gravitational lensing. In this
expression, $d_LH_0$ is expressed in units of kilometers per second.
These authors actually use the notation $\langle\log(dH_0)\rangle$ for
$a$ and $\pm$ for $\delta_a$.

In this paper, we work with the sample of \citet{tonryetal03}. This is not
the most up-to-date sample, but it is sufficient for our purpose. This
sample is comprised of 230 Type Ia SNe in the redshift range
$0<z<1.8$, with 79 of them being located at redshifts $z>0.3$ (including 5
at redshifts $z>0.9$). The values of $a$ and $\delta_a$ can be read directly
in the $8^{\rm th}$ and $9^{\rm th}$ columns of their Table~8, respectively.

\subsection{A Mock Catalog of Very-high $z$ Supernovae}

We generated a mock catalog of 43 SNe in the range $1.8<z<8.1$.
For each ``supernova,'' we need to choose a redshift $z$,
a measured value $a$, and an uncertainty $\delta_a$. There
is of course no rigorous method for doing that, since these SNe do not
exist. To provide a good coverage of the
redshift range of interest, we used 43 equally-spaced values of $z$ between
$z=1.8$ and $8.1$. To determine $\delta_a$, we first plotted $\delta_a$
versus $z$, for the \citet{tonryetal03} sample, to look for trends.
There is a large number of SNe with $z<0.1$, $\delta_a<0.05$. 
If we focus on the 79 SNe at redshift $z>0,3$,
we do not see any obvious trend, and in particular $\delta_a$ does not
appear to increase with redshift. So we chose, somewhat arbitrarily, the
9 SNe\footnote{We deliberately avoided sn97G and sn76cl, whose
uncertainties are much larger than those of any SNe at
comparable redshift.} at $z>0.828$. 
For these SNe, the mean and standard deviation of
the uncertainties are
$\bar{\delta}_a=0.0631$ and $\sigma_\delta=0.0113$, respectively. 
We then chose
the values of $\delta_a$ for our mock SNe randomly, by drawing them from
a normal distribution with mean $\bar{\delta}_a$ and
standard deviation $\sigma_\delta$. This ensures a smooth transition between 
the real and mock samples.

To determine $a$, we assume that the underlying cosmology
corresponds to a $\Lambda$CDM universe (as supported by the real sample).
We then use
\begin{equation}
a=a_\Lambda+\Delta a\,,
\end{equation}

\noindent where $a_\Lambda\equiv\log(d_\Lambda H_0)$ is the actual value of
$\log(d_LH_0)$ in a $\Lambda$CDM universe, and $\Delta a$ is a random number
drawn from a normal distribution with mean 0 and standard
deviation given by the value of $\delta_a$ we just calculated.

\section{THE EFFECT OF LENSING ON STATISTICS OF HIGH-Z SUPERNOVAE}

\subsection{Compounding the Errors}

As we explained in \S1, our goal is not to perform a rigorous error
analysis of the uncertainties resulting from the possibility of
lensing, but rather to estimate {\it a posteriori} the effect of
lensing on the uncertainties already present in the analysis.

We estimate the effect of lensing as follows: the distances of high-$z$
supernovae are reported in the literature in the format given by 
equation~(\ref{apmda}), where $\delta_a$ is the intrinsic uncertainty
(i.e. not caused by lensing). The distance $d_L$ is related to the
luminosity $L$ and flux $F$ by
\begin{equation}
F={L\over4\pi d_L^2}\,.
\label{flux}
\end{equation}

\noindent We use equations~(\ref{apmda}) and~(\ref{flux}) 
to eliminate $d_L$, and get
\begin{equation}
L^{1/2}H_0=10^a10^{\pm\delta_a}(4\pi F)^{1/2}\,.
\label{lsqh0}
\end{equation}

\noindent The effect of lensing will be to modify the flux $F$. To
account for it, we replace $F$ by $F\pm\Delta F$ in equation~(\ref{lsqh0}),
and perform a Taylor expansion
to first order in $\Delta F$,
\begin{equation}
L^{1/2}H_0=10^a10^{\pm\delta_a}(4\pi F)^{1/2}
\biggl(1\pm{\Delta F\over2F}\biggr)\,.
\end{equation}

\noindent
This expression reduces to
\begin{equation}
\log(d_LH_0)=a\pm\delta_a\pm{\Delta F\over2F\ln10}\,.
\label{errors1}
\end{equation}

\noindent The last term represents the uncertainty due to lensing. 
For a given supernova with magnification $\mu$, 
$\Delta F/F=\mu-1$. Of course, we will never
know the value of $\mu$ for a single source. But for a large number
of sources, we can use statistics. First, the simplest, lowest-order 
approximation for a ``typical'' value of $\mu$ is 
$\mu=\langle\mu\rangle\pm\sigma_\mu=1\pm\sigma_\mu$, or equivalently
$\Delta F/F\approx\sigma_\mu$. Equation~(\ref{errors1}) reduces to
\begin{equation}
\log(d_LH_0)=a\pm\delta_a\pm\delta_\mu\,,
\label{errors2}
\end{equation}

\begin{figure}
\begin{center}
\includegraphics[width=5in]{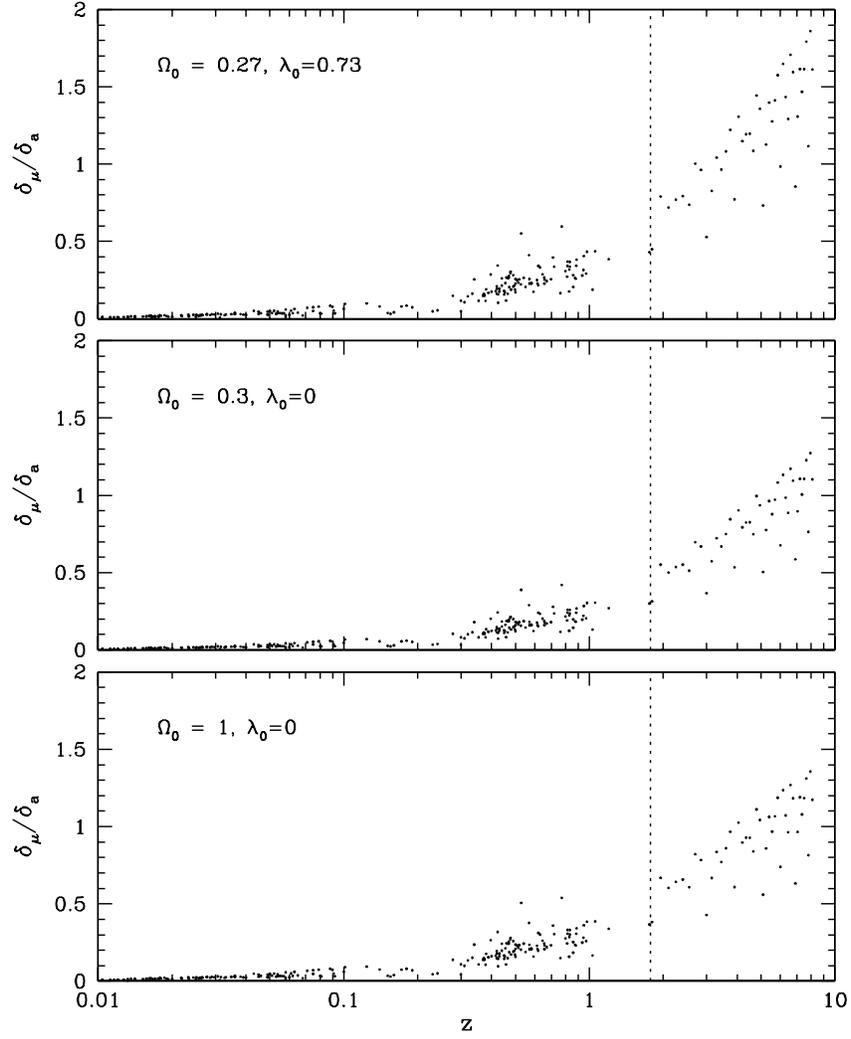}
\caption{Ratio $\delta_\mu/\delta_a$ versus redshift. The dotted
lines separate the real data of Tonry et al. (left side) from the
mock, high-redshift data (right side).
}
\label{ratio}
\end{center}
\end{figure}

\noindent where $\delta_\mu(z)=\sigma_\mu(z)/2\ln10$ can be computed
using the empirical relations plotted in Figure~\ref{sigma}. We
use the values of $a$ and $\delta_a$ reported by
\cite{tonryetal03} (their Table~8). In Figure~\ref{ratio},
we plot the ratio $\delta_\mu/\delta_a$ versus $z$ (left of the dotted lines).
This quantity increases with redshift, but never gets higher than 0.5 for
the Tonry et al. sample. Furthermore, we shall assume that $\delta_a$ and
$\delta_\mu$ are statistically independent, and combine them in quadrature,
using
\begin{equation}
\delta=(\delta_a^2+\delta_\mu^2)^{1/2}\,,
\end{equation}

\noindent where $\delta$ is the total error.
The contribution of lensing to this error is then of order 25\% at most.

For each supernova, we compute the quantity $\Delta(m-M)$ 
(deviation of the difference between
apparent and absolute magnitude, relative to an empty universe),
using
\begin{equation}
\Delta(m-M)=5\log(d_LH_0)-5\log(d_LH_0)_{\rm empty}
=5\log(d_LH_0)-5\log\Biggl[cz\biggl(1+{z\over2}\biggr)\Biggr]\,.
\end{equation}

\noindent We then average the quantities $\Delta(m-M)$ and $\delta$ 
in redshift bins, using

\begin{eqnarray}
\label{dmbin}
\left[\Delta(m-M)\right]_j&=&\Sigma_iw_i\Delta(m-M)/\Sigma_iw_i\,,\\
\label{deltabin}
\delta_j&=&(1/\Sigma_iw_i)^{1/2}\,,
\end{eqnarray}

\noindent where

\begin{equation}
\label{weights}
w_i=1/\delta_i^2\,,
\end{equation}

\noindent and the sums are over all data points $i$ in bin $j$
(note: eq.~[\ref{deltabin}] comes from $1/\delta_j^2=\sum_i[1/\delta_i^2]$).
Notice that this method of averaging is much fancier than what appears
to be done in the supernovae papers. For instance, Figure~9 of
\citet{tonryetal03} shows an averaging over redshift bins which is
based on the median of the data and apparently does not take into account
the uncertainties $\delta_a$ on the individual supernovae.

\begin{figure}
\begin{center}
\includegraphics[width=4.6in]{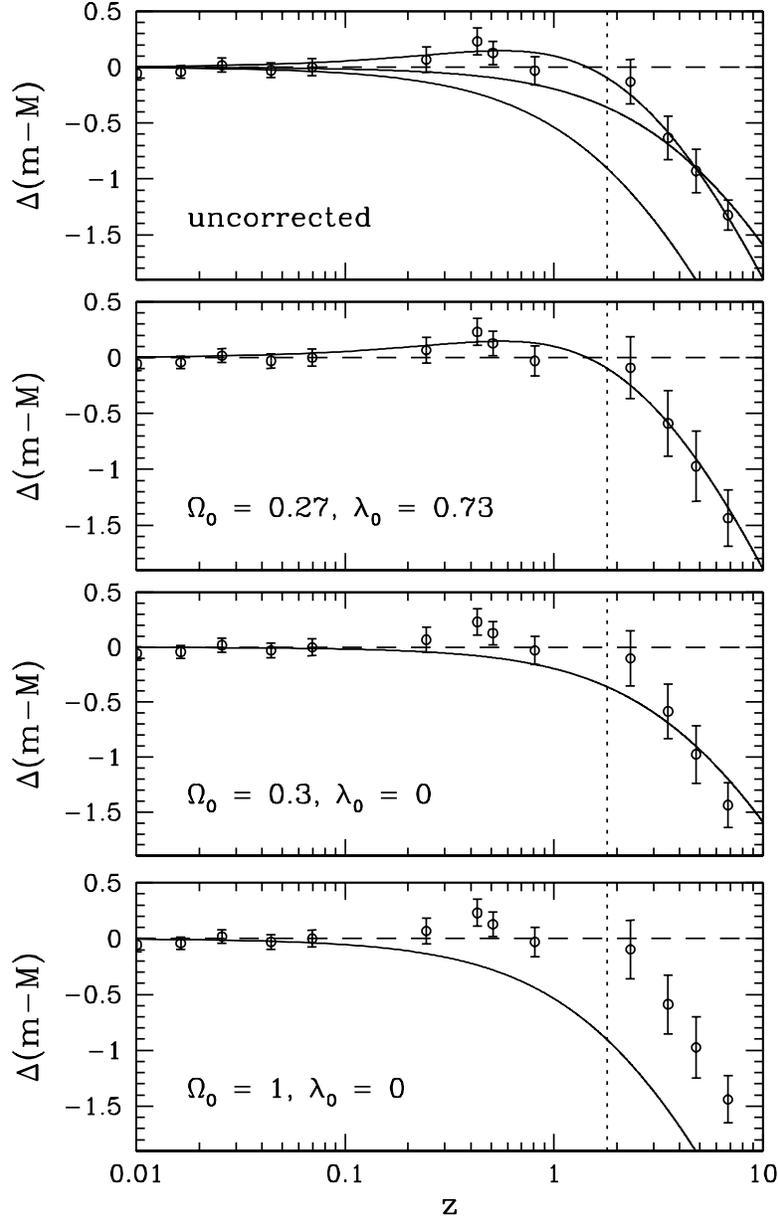}
\caption{Hubble diagram showing the magnitude deviation
$\Delta(m-M)$ relative
to an empty universe, for the three models considered.
The dotted lines separate the \citet{tonryetal03} data (on the left)
from the mock data (on the right).
In the top panel, the
three curves, from top to bottom, show the analytical result for the
cosmological models ($\Omega_0$,$\lambda_0$)=(0.27,0.73), (0.3,0.0),
and (1.0,0.0), respectively. The error bars show 90\% confidence level.
The last three panels reproduce the data of the
top panel, but have been corrected to account for lensing. Since this
correction is model-dependent, the three models are plotted on separate
panels.
}
\label{hubble}
\end{center}
\end{figure}

Figure~\ref{hubble} shows a {\it Hubble diagram}
[deviation $\Delta(m-M)$ versus redshift].
The data points and error bars on the left hand side of the dotted lines
correspond to the values given by 
equations~(\ref{dmbin}) and (\ref{deltabin}), respectively.
The three curves, from top to bottom, show the exact results for the
$\Lambda$CDM, low-density, and matter-dominated models, respectively.
The results support the flat $\Lambda$CDM model
and exclude the other models considered.
The other panels of Figure~\ref{hubble} show the effect of lensing (the
three models have to be plotted separately, because the correction due to
lensing, which uses the relations plotted in Fig.~\ref{sigma}, 
is model-dependent). 
This effect is totally negligible. The largest
correction to the error bars is about 10\% for the highest redshift bin,
for the $\Lambda$CDM model.

Clearly, the potential error introduced by lensing is negligible in
comparison to the intrinsic error in the measurement, at least for SNe at
redshifts $z<1.8$. \citet{gunnarssonetal06} and \citet{jonssonetal06}
reach the same conclusion,
We now estimate the effect of lensing on a
yet-undiscovered population of very-high-$z$ SNe, using our mock catalog. 
The ratios $\delta_\mu/\delta_a$ are plotted in Figure~\ref{ratio},
on the right hand side of the dotted lines.
The effect of lensing rapidly becomes important at redshift $z>2$,
especially for the $\Lambda$CDM models. We find many
SNe with $|\delta_\mu|>\delta_a$, that is, the correction due to lensing is
larger than the intrinsic uncertainty.

The points located on the right hand side of the dotted lines in
Figure~\ref{hubble} shows the results for the mock data.
The error bars get significantly bigger when lensing is
included. Furthermore, at redshift $z\approx3$, it becomes very difficult
to distinguish the open, low density model from the
cosmological constant model, because the theoretical curves intersect.
Keeping in mind the caveat that the mock catalog was built under the
assumption
that the underlying cosmology was $\Lambda$CDM, we see that the
Einstein-de~Sitter model is totally ruled out by SNe at $z>2$, but
the open model is not. Indeed, it is clear that SNe at $z>2$ would be
rather useless in distinguishing an open CDM and a $\Lambda$CDM model:
the theoretical curves get closer, whereas 
the error bars become larger.
It is, interestingly, the SNe in the redshift range $0.3<z<1$ that would
still provide the best discriminant between these two models, and data in
that redshift range are already available.

\subsection{Monte Carlo Approach}

The calculation presented in the previous section relies entirely
in the standard deviation $\sigma_\mu$ for estimating the uncertainties
caused by lensing. This approach would probably be sufficient if the
distributions of magnifications $P(\mu)$ were gaussian. However, 
for sources at large redshifts, $P(\mu)$ is strongly skewed, as Figure~1
shows. The large
majority of sources are demagnified, as the light reaching the observer 
travels mostly through underdense regions of the universe, while a
few sources are strongly magnified, especially those which happen to
be aligned with a massive galaxy at intermediate redshift.

\begin{figure}
\begin{center}
\includegraphics[width=5in]{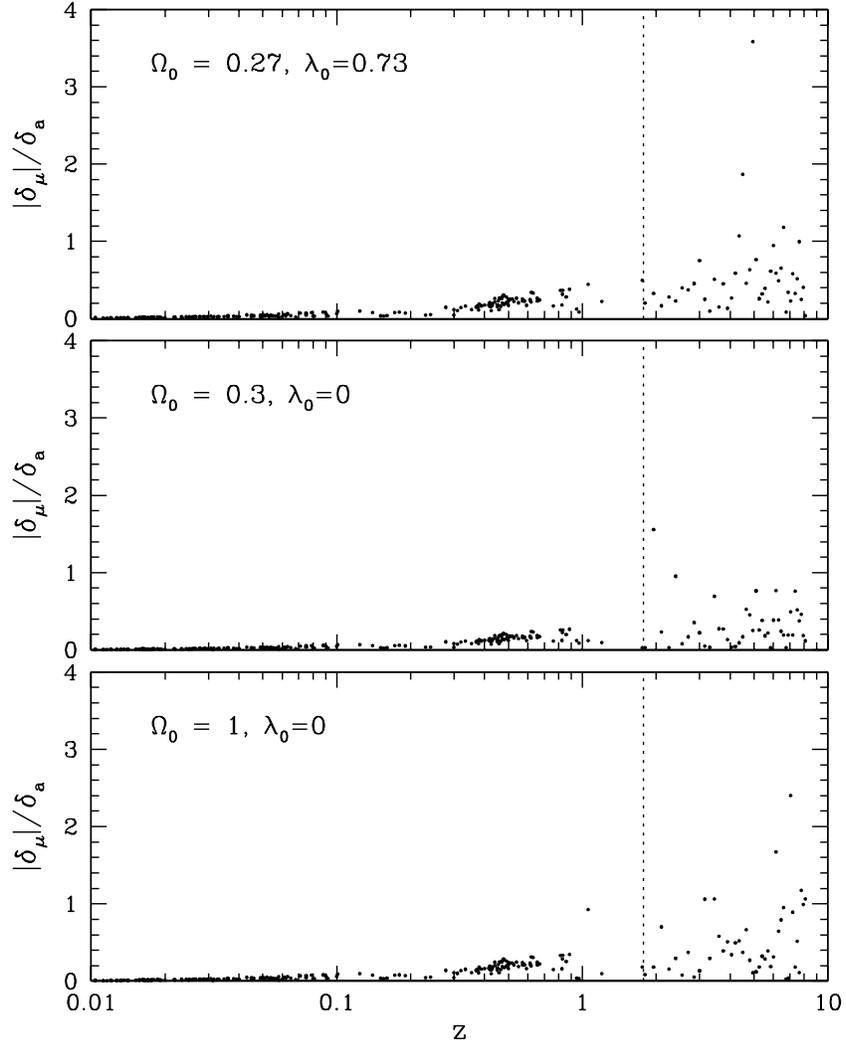}
\caption{Same as Figure \ref{ratio}, but with $\delta_\mu$
computed using the Monte Carlo approach for all SNe (real or mocked)
at redshift $z>0.9$.
}
\label{ratio2}
\end{center}
\end{figure}

To account for the distribution of magnifications, we consider all SNe
at redshifts $z>0.9$ (5 from the Tonry sample, 43 from the mock catalog).
For each one, we determine the distribution $P(\mu)$ at its redshift $z$, using
equation~(\ref{newpmu}), and then choose a
magnification $\mu$ by drawing it randomly from the distribution $P(\mu)$.
We then compute $\delta_\mu=\Delta F/2F\,\ln10=(\mu-1)/2\ln10$.
The resulting ratios $|\delta_\mu|/\delta_a$ are plotted in 
Figure~\ref{ratio2}.
Comparing with Figure~\ref{ratio}, we find only a few SNe for which
this ratio exceeds unity. For all redshifts and all models, we find that
the distributions $P(\mu)$
peak at a value $\mu_{\rm peak}<1$ such that 
$|\mu_{\rm peak}-1|<\sigma_\mu$. Hence, setting $\Delta F/F=\sigma_\mu$
(instead of $\Delta F/F=\mu-1$), as we did in \S4.1, overestimates the
effect of lensing for most SNe. However, the distributions are very skewed,
and as a result a few SNe are highly magnified, as Figure~\ref{ratio2}
shows.

\begin{figure}
\begin{center}
\includegraphics[width=4.6in]{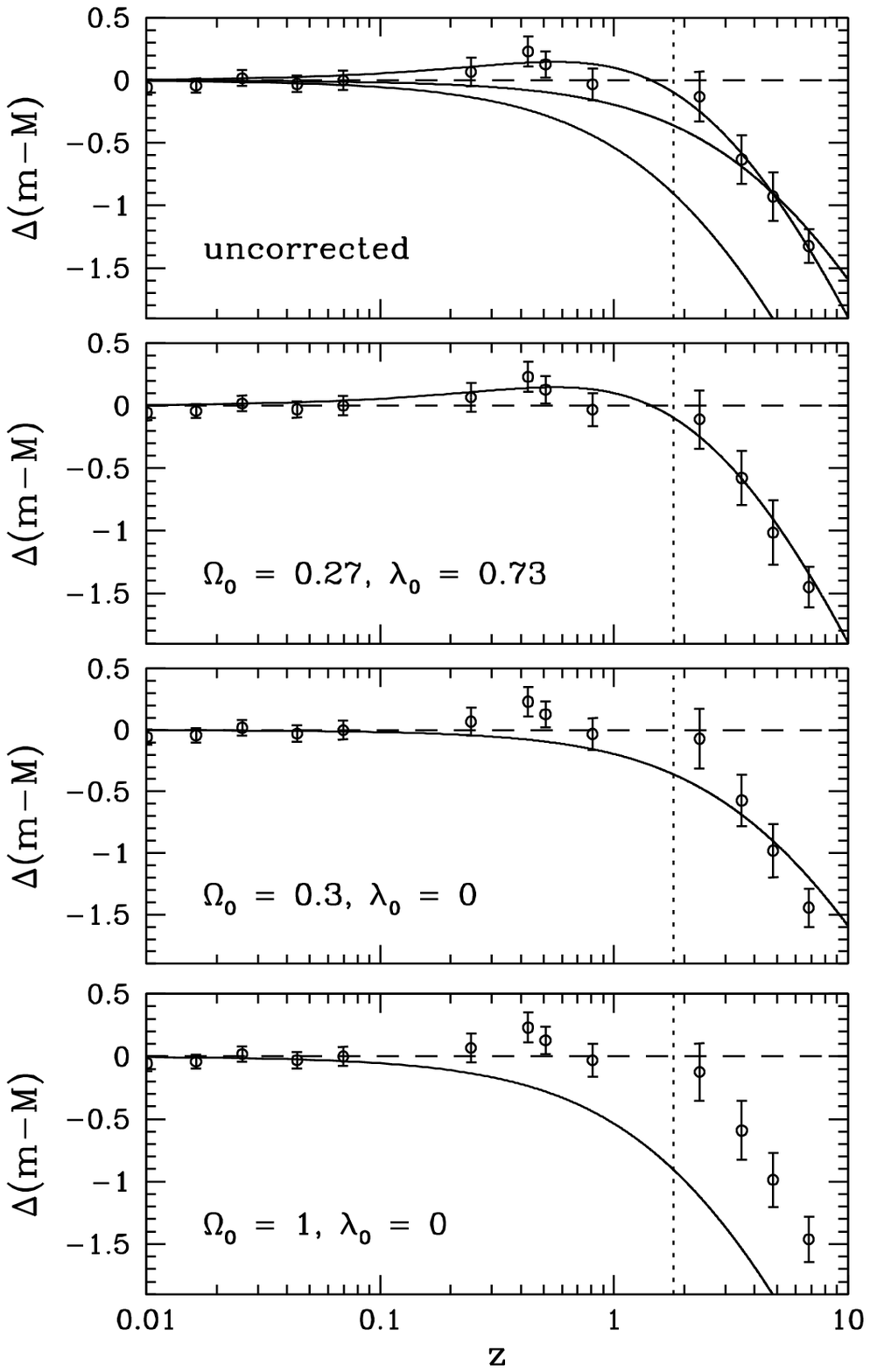}
\caption{Same as Figure \ref{hubble}, with errors bars computed
using the Monte Carlo approach.
}
\label{hubble_hz2}
\end{center}
\end{figure}

Figure~\ref{hubble_hz2} shows the resulting Hubble diagram. Comparing with
Figure~\ref{hubble}, we find that the error bars are significantly smaller.
The effect of lensing is less important when we use the actual distribution
of magnifications $P(\mu)$, and not only its standard deviation. However,
the results at $z>1.8$, which assume an underlying $\Lambda$CDM cosmology,
still cannot rule out the open CDM model; the error bars are still too large.

\section{DETECTION LIMIT AND BIASING}

So far, we have assumed that any supernova, with any value of $z$ and $\mu$,
can be observed. This assumption is probably valid over the range $0<z<2$,
which includes all the current observations. But as the redshift gets higher,
it becomes increasingly difficult to observe supernovae with current
and even future telescopes, because of the combined effect of the flux
reduction and the light being shifted to the near infrared. It might
just be impossible to detect a supernova at redshift $z>2$ unless, of
course, it is magnified by lensing. \citet{mf98} argue that, with
magnification taken into account, it might be possible to observe
Type~II SNe at redshifts up to $z=10$. Of course, if only the highly-magnified
SNe are observable, this introduces biasing, an effect that we must take
into account.

Here, we do not want to perform a detailed analysis similar to the one
of \citet{mf98}, but just to get a rough estimate of the importance
of biasing. The key results of the analysis of \citet{mf98} are
shown in their Figure~7, which shows, as a function of redshift,
the AB apparent magnitude in the J, K, L, and M bands, with and without
lensing. They also plot the expected flux limit of {\sl JWST\/}\footnote{They
called it {\sl NGST} back then.}. 
From this figure, we see that without lensing, the
apparent magnitudes are below the detection limit at high redshift. When
lensing is included, the magnitudes in the J and K bands are above the
detection limit. Their analysis was for Type~II SNe. Type~Ia SNe are
typically 1.5 magnitudes brighter, and therefore much easier to detect.
Neglecting the details of the spectra, we can simply take Figure~7
of \citet{mf98} and shift all the curves upward by 1.5 magnitudes.
We find that, without lensing, Type~Ia SNe would be visible in the
J and K band, and almost visible in the L band, at redshifts up to
$z\sim8$. With lensing, most SNe would be demagnified, but the reduction
in flux is typically of the order $10-20\%$, that is, a correction of
$0.103-0.198$ magnitudes. Hence, all high-$z$ Type~Ia SNe should be detectable,
using the proper telescope, and therefore we were justified to ignore any
biasing effect.
This being said, the identification of Type Ia SNe requires
that we obtain a spectrum, and this could be quite challenging
at these extreme redshifts.

We have assumed that the typical intrinsic uncertainties $\delta_a$
do not grow with redshift for $z>1.8$, based on the absence of
obvious trend at $z\lesssim1.8$. If the uncertainties do grow
with redshift, our conclusion that understanding the effect of
lensing at high redshift would be weakened, in the sense that
these data would be rather useless no matter how well lensing is
understood. Furthermore, it would reinforce our conclusion that
SNe at intermediate redshifts are more useful to discriminate between
different cosmological models.

We should also ask whether Type~Ia SNe at redshift $z=8$ can
actually exist. For a $\Lambda$CDM model with a Hubble constant of
$\rm 71\,km\,s^{-1}Mpc^{-1}$, this redshift corresponds to an age
of the universe
of $650\,\rm Myr$. Subtracting the formation and evolutionary time of
the progenitor, we are getting embarrassingly close to the big bang. Type~II
SNe would be a far better candidate for high-$z$ SNe, since the
evolutionary time of their progenitors are much shorter. But then only
the few that are magnified would be detectable, and their number might be
too small to perform any meaningful statistics on them.

\section{SUMMARY AND CONCLUSION}

We have performed a series of ray-tracing experiments using a
multiple lens-plane algorithm. We have determined the distributions
of  magnifications $P(\mu)$ for sources in the redshift range
$0<z<8$, for three different cosmological models. 
We have used these distributions to estimate
the effect of gravitational lensing on the determination of the
cosmological parameters with high-redshift Type~Ia
supernovae. We used a generic, {\it a posteriori} approach which is
not tied to any particular sample.

We found that errors introduced by lensing are unimportant for SNe
with redshift $z<1.8$. These errors are negligible compared to the
intrinsic errors already present in the supernovae data. Since those
intrinsic errors do not prevent us from determining the cosmological
parameters, the additional errors introduced by lensing have no consequences.
A similar conclusion was reached by \citet{alderingetal06}.

Using a mock catalog of high-$z$ SNe, extending to $z=8.1$, we showed that
the effect of lensing on a hypothetical
population of SNe at redshifts $z>2$ could be quite significant, and
must be understood before such SNe could be used to constrain cosmological
models. Furthermore, the open CDM and $\Lambda$CDM are difficult to distinguish
at that redshift. We showed that, even if SNe at redshift $z\sim8$ were
ever discovered, it is the SNe in the range $z=0.3-1$ that would still
provide the best discriminant between these two models. The data at that
redshift already exist, and they support the $\Lambda$CDM model.

\acknowledgments
This work benefited from stimulating discussions with Gilbert Holder,
Daniel Holz, Eric Linder, Massimo Meneghetti, and Christopher Vale.
The calculations were performed at the Texas 
Advanced Computing Center, University of Texas, and the
Laboratoire d'Astrophysique Num\'erique, Universit\'e Laval.
HM thanks the Canada Research Chair program and NSERC for
support. PP thanks Uro\v s Seljak for various advice and hospitality during
the fruitful visit at the Abdus Salam ICTP in Trieste.

%


\end{document}